\documentclass[journal]{IEEEtran}
\usepackage{mathtools}
\usepackage{amsmath}
\usepackage{breqn}
\usepackage{tabularx}
\usepackage{tabu}
\usepackage{booktabs}
\usepackage{tikz}
\usepackage{subfig}
\usetikzlibrary{shapes,arrows,positioning,calc}
\usepackage[noadjust]{cite}
\usepackage{csquotes}
\usepackage{dirtytalk}
\usepackage{amsfonts}
\usetikzlibrary{decorations.pathmorphing}

\DeclarePairedDelimiter{\diagfences}{(}{)}
\newcommand{\diag}{\operatorname{diag}\diagfences}

\DeclareFontFamily{U}{mathb}{\hyphenchar\font45}
\DeclareFontShape{U}{mathb}{m}{n}{<5> <6> <7> <8> <9> <10> gen * mathb
<10.95> mathb10 <12> <14.4> <17.28> <20.74> <24.88> mathb12}{}
\DeclareSymbolFont{mathb}{U}{mathb}{m}{n}

\DeclareMathSymbol{\rcirclearrow}{\mathbin}{mathb}{'367}
\makeatletter
\newcommand*{\Rcirclearrow}{\mathpalette\@Rcirclearrow{}}
\newcommand*{\@Rcirclearrow}[1]{%
    \mathbin{\ooalign{\hphantom{$#1\rcirclearrow$}\cr\hss\raisebox{0.05ex}{%
                \scalebox{0.9}{%
                    \rotatebox[origin=c]{270}{%
                        $#1\rcirclearrow$}}}\hss}}}
\makeatother

\ifCLASSINFOpdf
\else
\fi
\usepackage{fixltx2e}

\usepackage{stfloats}
\begin{document}

\tikzset{
block/.style = {draw, fill=white, rectangle, minimum height=2.5em, minimum width=2.5em},
block2/.style = {draw, fill=white, rectangle, minimum height=1.5cm, minimum width=1cm},
lblock/.style = {draw, fill=white, rectangle, minimum height=8em, minimum width=3em},
tmp/.style  = {coordinate}, 
point/.style  = {coordinate},
sum/.style= {draw, fill=white, circle, node distance=1cm},
input/.style = {coordinate},
output/.style= {coordinate},
pinstyle/.style = {pin edge={to-,thin,black}},
blockbus/.style = {draw, fill=black, rectangle, minimum height=25em, minimum width=0.0em,scale=0.1},
generator/.style = {draw, fill=white, circle, minimum height=0.5cm, minimum width=0.5cm},
tr/.style= {draw, fill=white, fill opacity = 0, circle, minimum height=0.5cm},
load/.style     = {draw, isosceles triangle, rotate=30, minimum height = 1em, isosceles triangle apex angle=60},
capacitor/.style ={minimum width = 0.5cm,minimum height = 0.2cm,
        draw=none,
        append after command={
            [shorten <= -0.5\pgflinewidth]
            ([shift={(-1.5\pgflinewidth,-0.5\pgflinewidth)}]\tikzlastnode.north east)
        edge([shift={( 0.5\pgflinewidth,-0.5\pgflinewidth)}]\tikzlastnode.north west) 
            ([shift={( 0.5\pgflinewidth,+0.5\pgflinewidth)}]\tikzlastnode.south west)
        edge([shift={(-1.0\pgflinewidth,+0.5\pgflinewidth)}]\tikzlastnode.south east)
        }
    }
}
\tikzset{snake it/.style={decorate, decoration=snake}}
\title{Nonlinear Model Predictive Control \\ of Variable Speed Hydropower \\ for Provision of Fast Frequency Reserves}
%
%
%

\author{Tor~Inge~Reigstad,~
        Kjetil~Uhlen
\thanks{This work was supported by the Research Council of Norway under Grant 257588 and by the Norwegian  Research  Centre for Hydropower Technology (HydroCen).}
\thanks{T.I. Reigstad and K. Uhlen are with the Department for Electric Power Engineering, Norwegian University of Science and Technology (NTNU), NO-7491 Trondheim, Norway (email: tor.inge.reigstad@ntnu.no, kjetil.uhlen@ntnu.no)}
}

\maketitle

\begin{abstract}

This paper presents the development of a non-linear model predictive controller (MPC) for controlling variable speed hydropower (VSHP) plants. The MPC coordinates the turbine controller with the virtual synchronous generator (VSG) control of the power electronics converter to optimize the plant's performance. The main objective is to deliver a fast power response to frequency deviations by utilizing the kinetic energy of the turbine and generator. This is made possible by allowing the turbine rotational speed to deviate temporarily from its optimal speed. In addition, the efficiency should be maximized while keeping the electric and hydraulic variables within their constraints. The simulation results show that the proposed MPC is also able to damp power oscillations in the grid, reduce water hammering in the penstock and improve the future estimation of turbine head, turbine power and turbine flow. This ensures that the turbine head does not exceed its limits and that the overshoot in the turbine speed after a disturbance is reduced. Besides, the VSG converter control enables a fast power response by utilizing the rotational energy of the turbine and generator. Thereby, the VSHP can provide a significant amount of fast frequency reserves (FFR) to the grid.

\end{abstract}


%
\IEEEpeerreviewmaketitle

\section{Introduction}
%
%
%
%

\IEEEPARstart{H}{ydropower} is an important contributor in providing power system flexibility and will remain a significant source of large-scale energy storage in the future \cite{graabak2019balancing}. The share of variable renewable energy, such as wind and solar power, is increasing. Thus, more flexible generation and demand are required to control the balance of the grid and to maintain power system security. Variable speed operation of hydropower plants has the potential to provide faster control of active and reactive power than conventional hydropower plants. This is achieved by applying frequency converter technology and implementation of virtual inertia (VI) control by utilizing the kinetic energy stored in the turbine and generator. The hypothesis is that the variable speed hydropower (VSHP) can offer additional ancillary services, contributing to improved frequency control and maintaining the grid stability, allowing for higher penetration of variable renewables in the grid. A robust and well-functioning control system needs to be developed to coordinate the control of the hydraulic system and the converter control. The control system must optimize the operation of the power plant while considering the constraints in the electric and the hydraulic systems to maximize the potential of the kinetic energy.

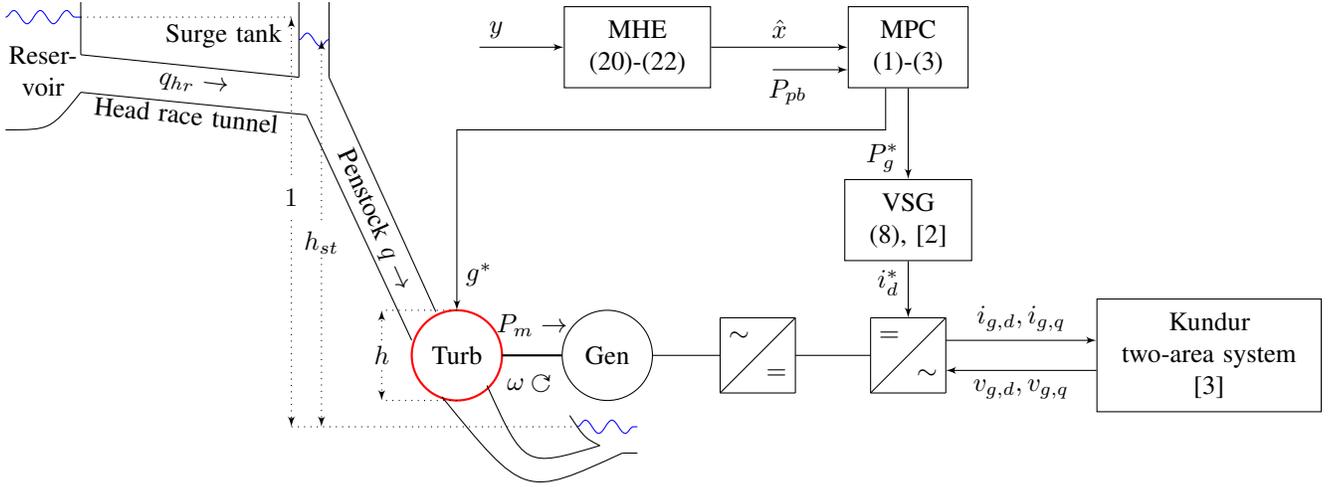
\begin{figure*}[!t]
\centering
\begin{tikzpicture}[auto, node distance=1cm,>=latex']
    \path [draw=blue,snake it]
    (-1,0.5) -- (0,0.5) ;
    \path [draw=blue,snake it]
    (2.9,0.2) -- (3.3,0.2) ;
    \draw(0,0.7)--(0,0);
    \draw(-1,-1) .. controls (-0.4,-1) .. (0,-0.5);
    \draw(0,0)--(2.9,-0.3);
    \draw(0,-0.5)--(3,-0.8);
    \draw(3,-0.8)--(4.4,-3.8);
    \draw(3.3,-0.3)--(4.72,-3.42);
    \draw(2.9,-0.3)--(2.9,0.7);
    \draw(3.3,-0.3)--(3.3,0.7);
    \draw[red,thick](5,-4) circle (0.6);
    \draw(5.4,-4.4) .. controls (5.9,-5.5) .. (6.9,-5.2);
    \draw(4.8,-4.55) .. controls (5.9,-5.9) .. (7.2,-5.3);
    \draw(6.5,-4.8) .. controls (6.7,-5.1) .. (6.9,-5.2);
    \draw(7.2,-5.3)--(7.4,-5.3);
    \path [draw=blue,snake it]
    (6.6,-4.95) -- (7.4,-4.95) ;
    \node [align=center,rotate=-7] at (1.5,-0.4) {$q_{hr}\rightarrow$};
    \node [align=center,rotate=-6] at (1.4,-0.80) {Head race tunnel};
    \node [align=center,rotate=-66] at (3.9,-2.2) {Penstock $q\rightarrow$};
    \node [align=center] at (5,-4) {Turb};
    \node [align=center] at (-0.5,-0.25) {\begin{tabular}{c} Reser-\\ voir\\ \end{tabular}};
    \node [align=center] at (1.9,0.25) {Surge tank};
    
    \node [align=center] at (3.2,-2.5) {$h_{st}$};
    \draw[->,dotted](3.2,-2.2)--(3.2,0.2);
    \draw[->,dotted](3.2,-2.7)--(3.2,-4.95);
    
    \node [align=center] at (2.8,-1.9) {$1$};
    \draw[->,dotted](2.8,-1.6)--(2.8,0.5);
    \draw[->,dotted](2.8,-2.2)--(2.8,-4.95);
    
    \draw[dotted](0,0.5)--(2.8,0.5);
    \draw[dotted](2.8,-4.95)--(6.6,-4.95);
    
    \node [align=center] at (4,-4) {$h$};
    \draw[->,dotted](4,-3.8)--(4,-3.4);
    \draw[->,dotted](4,-4.2)--(4,-4.6);
    
    \draw[dotted](5,-3.4)--(4,-3.4);
    \draw[dotted](5,-4.6)--(4,-4.6);
    
    \draw(7.0,-4) circle (0.6);
    \node [align=center] at (7.0,-4) {Gen};
    
    \draw[thick](5.6,-4)--(6.4,-4);
    
    \draw(8.5,-3.5)--(9.5,-3.5);
    \draw(8.5,-4.5)--(9.5,-4.5);
    \draw(8.5,-3.5)--(8.5,-4.5);
    \draw(9.5,-3.5)--(9.5,-4.5);
    \draw(8.5,-4.5)--(9.5,-3.5);
    \node [align=center] at (8.75,-3.75) {$\sim$};
    \node [align=center] at (9.25,-4.25) {$=$};
    
    \draw(7.6,-4)--(8.5,-4);
    
    \draw(10.5,-3.5)--(11.5,-3.5);
    \draw(10.5,-4.5)--(11.5,-4.5);
    \draw(10.5,-3.5)--(10.5,-4.5);
    \draw(11.5,-3.5)--(11.5,-4.5);
    \draw(10.5,-4.5)--(11.5,-3.5);
    \node [align=center] at (10.75,-3.75) {$=$};
    \node [align=center] at (11.25,-4.25) {$\sim$};
    
    \draw(9.5,-4)--(10.5,-4);
    

    
    \node [align=center] at (6.0,-3.6) {$P_{m}\rightarrow$};
    \node [align=center] at (6.0,-4.4) {$\omega \Rcirclearrow$};
    
    \node [point] (start) {};
    \node [point, right of=start, node distance = 11.0cm] (startr) {};
    \node [block, above of=startr, node distance = 0.1cm] (MPC) {\begin{tabular}{c}
         MPC \\
         \eqref{objfunk}-\eqref{posdef}\\
    \end{tabular}};
    \node [point, below of=MPC, node distance = 2.3cm] (MPCb) {};
    \node [block, left of=MPCb, node distance = 0.0cm] (VSG) {\begin{tabular}{c}
         VSG \\
         \eqref{eqvsg}, \cite{reigstad2020variable}\\
    \end{tabular}};
    \node [point, below of=MPC, node distance = 1.4cm] (MPCb2) {};
    \draw [->] (MPC) -- (MPCb2) -| node[anchor=east,pos=0.6]{$P_g^*$} (VSG);
    \node [point, below of=MPC, node distance = 3.2cm] (MPCb3) {};
    \node [point, below of=MPC, node distance = 3.6cm] (MPCb4) {};
    \draw [->] (VSG) |- (MPCb3) node[anchor=east,pos=0.4]{$i_d^*$} -| (MPCb4);
    \node [block, left of=MPC, node distance = 3.6cm] (MHE) {\begin{tabular}{c}
         MHE \\
         \eqref{eqcostmhe}-\eqref{eqcostmhe3}\\
    \end{tabular}};
    \draw [->] (MHE) -- node[anchor=south,pos=0.5]{$\hat{x}$} (MPC);
    \node [point, left of=MHE, node distance = 2.1cm] (MHEl) {};
    \draw [->] (MHEl) -- node[anchor=south,pos=0.2]{$y$} (MHE);
    
    \node [point, below of=MPC, node distance = 4.1cm] (MPCb5) {};
    \node [block, right of=MPCb5, node distance = 4.0cm] (grid) {\begin{tabular}{c}
         Kundur \\
         two-area system\\
         \cite{reigstad2019variable} \\
        \end{tabular}};
    \node [point, below of=MPC, node distance = 3.9cm] (MPCb6) {};
    \node [point, below of=MPC, node distance = 4.3cm] (MPCb7) {};
    \node [point, right of=MPCb6, node distance = 0.5cm] (MPCb6r) {};
    \node [point, right of=MPCb7, node distance = 0.5cm] (MPCb7r) {};
    \draw [->] (MPCb6r) -- node[anchor=south,pos=0.5]{$i_{g,d},i_{g,q}$} ([yshift=0.2cm]grid.west);
    \draw [->] ([yshift=-0.2cm]grid.west) -- node[anchor=north,pos=0.5]{$v_{g,d},v_{g,q}$} (MPCb7r) ;
    

    
    \node [point, below of=MHEl, node distance = 0.7cm] (MHElb) {};
    \node [point, left of=MPC, node distance = 1.8cm] (MPCl) {};
    \node [point, below of=MPCl, node distance = 0.3cm] (MPClb) {};
    \draw [->]  (MPClb) -- node[anchor=north,pos=0.2]{$P_{pb}$} ([yshift=-0.3cm]MPC.west);
    
    \draw [->] ([xshift=-0.3cm]MPC.south) |-  (5,-1.0) -- node[anchor=west,pos=0.8]{$g^*$} (5,-3.4);

\end{tikzpicture}
\caption{VSHP plant layout with equations for modelling and control structure} \label{figvshp}
\end{figure*}

This paper aims to utilize nonlinear model predictive control (MPC) for controlling a VSHP plant. The main goal is to optimize the frequency support capabilities of the power plant while keeping the electric and hydraulic variables within their limits. One of the benefits by variable speed hydropower plant is that the kinetic energy of the turbine and generator can be utilized immediately to provide fast frequency reserves (FFR) to stabilize the grid. To maximize the provision of frequency support services and the efficiency of the system, the control of the turbine should be as fast as possible without exceeding the constraints given by the hydraulic system. A conventional control system with PID-controllers will become slow because it has to be tuned cautiously to avoid exceeding these constraints and thereby causing damage to the system. Therefore, an advanced control system is developed to optimize the system with the given constraints. A virtual synchronous generator (VSG) controls the VSHP output power to reduce frequency deviation while the MPC coordinates the VSG control and the control of the turbine, as shown in Figure \ref{figvshp}. The MPC also assures that both the electrical and hydraulic constraints are fulfilled. At the same time, the MPC limits the deviation of the turbine rotational speed to maximize the efficiency of the system.

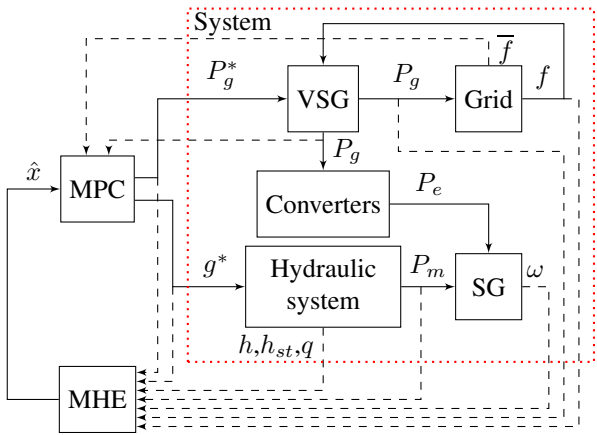
\begin{figure}
    \centering
    \begin{tikzpicture}[auto, node distance=2.5cm,>=latex']
    
    \node [block] (mpc) {MPC};
    \node [point, above of =mpc, node distance = 1.2cm] (mpca) {};
    \node [point, below of =mpc, node distance = 1.3cm] (mpcb) {};
    \node [point, right of =mpca, node distance = 0.8cm] (mpcar) {};
    \node [point, right of =mpcb, node distance = 1.0cm] (mpcbr) {};

    \node [block, right of=mpca, node distance = 3.0cm] (VSG) {VSG};
    \draw [->] ([yshift=0.15cm]mpc.east) -| (mpcar) -- node[anchor=south,pos=0.5]{$P_g^*$} (VSG);
    
    \node [block, right of=mpcb, node distance = 3.0cm] (hm) {\begin{tabular}{c}
         Hydraulic \\
         system
    \end{tabular}};
    \draw [->] ([yshift=-0.15cm]mpc.east) -| (mpcbr) -- node[anchor=south,pos=0.6]{$g^*$} (hm);
    
    \node [block, right of=VSG, node distance = 2.2cm] (gm) {Grid};
    \draw [->] (VSG) -- node[anchor=south,pos=0.5]{$P_g$} (gm);
    
    \node [block, below of=VSG, node distance = 1.4cm] (conv) {Converters};
    \draw [->] (VSG) -- node[anchor=west,pos=0.5]{$P_g$} (conv);

    \node [block, right of=hm, node distance = 2.2cm] (sg) {SG};
    \draw [->] (conv) -| node[anchor=south,pos=0.2]{$P_e$} (sg);
    
    \node [block, below of=mpc, node distance = 2.8cm] (mhe) {MHE};
    \draw [->] (hm) -- node[anchor=south,pos=0.5]{$P_m$} (sg);
    
    \node [point, below of =mpcar, node distance = 1.05cm] (mpcarb) {};

    \draw [->,dashed] (mpcarb) |-  ([yshift=0.36cm]mhe.east);
    \draw [->,dashed] (mpcbr) |-  ([yshift=0.24cm]mhe.east);
    \draw [->,dashed] (hm) |- node[anchor=east,pos=0.15]{$h$,$h_{st}$,$q$} ([yshift=0.12cm]mhe.east);
    \node [point, right of =hm, node distance = 1.3cm] (hmr) {};
    \draw [->,dashed] (hmr) |-  ([yshift=0.0cm]mhe.east);
    \node [point, right of =sg, node distance = 0.8cm] (sgr) {};
    \draw [->,dashed] (sg) -- node[anchor=south,pos=0.5]{$\omega$}(sgr) |-  ([yshift=-0.12cm]mhe.east);
    \node [point, right of =VSG, node distance = 1.0cm] (vsgr) {};
    \node [point, below of =vsgr, node distance = 0.7cm] (vsgrb) {};
    \node [point, right of =vsgrb, node distance = 2.2cm] (vsgrbr) {};
    \draw [->,dashed] (vsgr) -- (vsgrb) -- (vsgrbr) |-  ([yshift=-0.24cm]mhe.east);
    
    \node [point, right of =gm, node distance = 1.0cm] (gmr) {};
    \node [point, right of =gm, node distance = 1.2cm] (gmrr) {};
    \node [point, above of =gmr, node distance = 1.0cm] (gmra) {};
    \draw [->] (gm) -- node[anchor=south,pos=0.5]{$f$} (gmr) -- (gmra) -| (VSG);
    \draw [->,dashed] (gmr) -- (gmrr) |- ([yshift=-0.36cm]mhe.east);
    
    \node [point, left of =mpc, node distance = 1.2cm] (mpcl) {};
    \draw [->] (mhe) -| (mpcl) -- node[anchor=south,pos=0.5]{$\hat{x}$} (mpc);
    
    \node [point, above of =gm, node distance = 0.8cm] (gma) {};
    \draw [->,dashed] (gm) -- node[anchor=west,pos=0.6]{$\overline{f}$} (gma) -| ([xshift=-0.15cm]mpc.north);
    \node [point, below of =VSG, node distance = 0.55cm] (vsgb) {};
    \draw [->,dashed] (vsgb) -|  ([xshift=0.15cm]mpc.north);
    
    \draw[red,thick,dotted]  ($(hm)+(-1.8,-1.0)$)  rectangle ($(gm)+(1.4,1.2)$) ;
    \node [align=left] at ($(VSG)+(-1.2,1.0)$) {System};
    
\end{tikzpicture}
\caption{Control layout of VSHP plant with MPC and MHE} \label{fig:vshp}
\end{figure}

The MPC of the proposed control scheme calculates and supplies the optimal VSHP output power reference $P_g^*$ to the VSG and guide vane opening reference $g^*$ to the turbine, as indicated in Figure \ref{fig:vshp}. As the VSG controls the VSHP output power $P_g$ with the frequency $f$ as input, there is a direct relationship between these variables, as in a conventional hydropower plant. However, due to the converter technology, the turbine rotational speed does not need to follow the frequency. Thereby, the VSHP output power can be controlled quicker by utilizing the rotational energy of the turbine and generator. The ancillary service capabilities are therefore no longer limited by the slow governor, as in a conventional hydropower plant. This opens new possibilities such as faster frequency control and other grid ancillary services, but it also necessitates proper co-ordination of the controls.

The nonlinear MPC controller is based on the VSHP models presented in \cite{reigstad2019modelling} and \cite{reigstad2019variable}, and is combined with the VSG control approach presented in \cite{reigstad2020variable}. The controller is a further development of the linear MPC for VSHP presented in \cite{reigstad2020optimized}. The motivation for using non-linear MPC control is to achieve more accurate results of the optimization problem and better performance of the controller. The MPC is also improved with models and control functions for damping of pressure waves in the penstock and damping of low-frequency power oscillations in the grid. Moving horizon estimation (MHE) is utilized as an observer model instead of a Kalman filter. 

MPC control systems for control of conventional hydropower governors have previously been investigated, however, the research is limited. These works assume direct-connected generators, such that the turbine rotational speed is following the grid frequency. Therefore, the hydraulic models do not consider varying turbine rotational speed. Additionally, they do not optimize control with regards to the provision of frequency services. A local MPC controller is developed in \cite{beus2018application} for turbine governor control. This work uses a simple MPC dynamic model of the system, including a governor with limits on the speed of the guide vane opening and a linearized (HYGOV) model for representing a Francis turbine. A more detailed model of the hydraulic system is utilized in \cite{zhang2015nonlinear} where a non-linear predictive control system is presented. The control system includes a terminal penalty function that proves Lyapunov stability for the discrete system. In \cite{kishor2007nonlinear}, the guide vane opening is controlled by a neural network-based nonlinear predictive controller to optimize the control of the turbine power. A multi-mode MPC scheme is proposed in \cite{zheng2016design} for excitation control and load scheduling of a hydropower plant. Experimental results indicate both increase performance of voltage regulating, damping and control of the turbine governor.

MPC has also been utilized for frequency control, as presented in \cite{elsisi2018improving}. Here, the MPC design for load frequency control of superconducting magnetic storage and capacitive energy storage is optimized. Load frequency control by MPC is studied in \cite{ersdal2013applying,ersdal2016model,ersdal2016model2}, where both linear and nonlinear centralized MPC solutions take into account limitations on tie-line power flow, generation capacity, and generation rate of change. Another possibility is to utilize MPC for damping of low damped electromechanical modes by minimizing the generator's frequency deviation from the average system frequency with the use of a global MPC-based grid controller \cite{fuchs2014stabilization,sanz2017effective, azad2013damping, jain2015model}. Looking beyond frequency control, similar control layouts can be applied to control voltage and ensure voltage stability \cite{imhof2014voltage}.

Although MPC based control systems have been proposed for both conventional hydropower plant, frequency control and damping of power oscillations, little or no work embrace VSHP and optimization of frequency support capabilities by utilizing the kinetic energy of the turbine and generator. This paper contributes to further development of the concept proposed in \cite{reigstad2020optimized} by improving its efficiency and accuracy and adding functionalities like power oscillation damping and modelling of water hammering in the penstock.

The paper is organized as follows: The MPC theory, the control objectives for the MPC controller and the development of the MPC model are presented in Section \ref{MPC} while the MHE is presented in Section \ref{sec:mhe}. The results and discussions are given in Section \ref{Results} and the conclusion in Section \ref{Conclusion}.

\section{Model Predictive Control}\label{MPC}

MPC is a well-developed and widely used method in process control, offering great advantages compared to traditional PID-controllers. By utilizing dynamic models of the process to solve an optimization problem, the MPCs handle both constraints, nonlinear systems and multiple-input, multiple-output (MIMO) systems. MPCs are more robust and may offer a faster and smoother response and lower rising time, settling time and overshoots compared to PID-controllers. MPC is a closed-loop optimization problem where a discrete-time model is optimized on a time horizon from $t=0$ to $t=N$.

The principle of model predictive control (MPC) is formulated in \cite{mayne2000constrained}:

\begin{displayquote}
Model predictive control is a form of control in which the current control action is obtained by solving, at each sampling instant, a finite horizon open loop optimal control problem, using the current state of the plant as the initial state; the optimization yields an optimal control sequence and the first control in this sequence is applied to the plant.
\end{displayquote}

A nonlinear MPC model with a quadratic objective function, nonlinear equality constraints and linear inequality constraints is used in this paper. The model \eqref{objfunk}-\eqref{posdef} includes the cost for the error of state/variables values $x$, changes in state values $\Delta x$, the error of input values $u$, changes in input values $\Delta u$ and the cost for exceeding the limitations on the states with the use of slack variables $\epsilon$.

\begin{multline}\label{objfunk}
    \min_{x \in \mathbb{R}^n, u \in \mathbb{R}^m} f(x,u) = \sum_{t=0}^{N-1} \frac{1}{2} x_{t+1}^{\text{T}}Q_{t+1}x_{t+1}  \\
    + d_{xt+1} x_{t+1} + \frac{1}{2} \Delta x_{t+1}^{\text{T}} Q_{\Delta t} \Delta x_{t+1}+ \frac{1}{2} u_{t}^{\text{T}}R_{t}u_{t}  \\ 
    + d_{ut} u_{t} + \frac{1}{2} \Delta u_{t}^{\text{T}} R_{\Delta t} \Delta u_{t} + \rho^{\text{T}} \epsilon_t + \frac{1}{2} \epsilon_t^{\text{T}}S \epsilon_t
\end{multline}

subjected to

\begin{equation}\label{constr}
    \begin{aligned}
        x_{t+1} &= g\left( x_t, u_t \right) \\
        x_0, u_{-1} &= \text{given} \\
        x^{\text{low}} - \epsilon &\leq x_t \leq x^{\text{high}} +\epsilon \\
        -\Delta x^{\text{high}} &\leq \Delta x_t \leq \Delta x^{\text{high}}\\
        A_{ineq} x_t &+ B_{ineq} u_t \leq b_{ineq} \\
        u^{\text{low}} &\leq u_t \leq u^{\text{high}} \\
        -\Delta u^{\text{high}} &\leq \Delta u_t \leq \Delta u^{\text{high}} \\
    \end{aligned}
    \quad
    \begin{aligned}
        t &= 0, \dots, N-1 \\
        \\
        t &= 1, \dots, N\\
        t &= 1, \dots, N\\
        t &= 1, \dots, N\\
        t &= 0, \dots, N-1\\
        t &= 0, \dots, N-1\\
    \end{aligned}
\end{equation}

where

\begin{equation}\label{posdef}
    \begin{aligned}
        Q_t &\succeq 0 \\
        Q_{\Delta t} &\succeq 0 \\
        R_t &\succeq 0 \\
        R_{\Delta t} &\succeq 0 \\
        \Delta x_t &= x_t -x_{t-1} \\
        \Delta u_t &= u_t -u_{t-1} \\
        \epsilon &\in \mathbb{R}^n_x \geq 0 \\
        \rho &\in \mathbb{R}^n_x \geq 0 \\
        S &\in \text{diag} \left\{s_1, \dots, s_{n_x} \right\}, s_i \geq 0,
    \end{aligned}
    \begin{aligned}
        t &= 1, \dots, N\\
        t &= 1, \dots, N\\
        t &= 0, \dots, N-1\\
        t &= 0, \dots, N-1\\
        t &= 1, \dots, N\\
        t &= 0, \dots, N-1\\
        \\
        \\
          i&=\left\{1,\dots,n_x\right\}
    \end{aligned}
\end{equation}

The parameters of the MPC functions \eqref{objfunk}-\eqref{posdef} are derived from the MPC dynamic model given in Section \ref{sec:mpcdyn}, the costs defined in Section \ref{ch:costs} and the constraints and slack variables presented in Section \ref{ch:limits}. Solution of the nonlinear optimization problem for each time step is found by CasAdi \cite{andersson2019casadi} in MATLAB, using the direct multiple shooting method and the IPOPT solver \cite{wachter2006implementation}.

\subsection{Control Objectives of the MPC Controller}\label{ch:objectives}

The control objectives of the MPC are an extended version of those presented in \cite{reigstad2020optimized}.

\begin{itemize}
    \item Primary frequency control:
    \begin{itemize}
        \item Provide power reference $P_g^*$ to VSG.
        \item Minimize deviation in grid frequency $\Delta f$.
        \item Keep the converter power $P_g$ within its limits.
        \item Power oscillation damper (POD).
    \end{itemize}
    \item Hydraulic system control:
    \begin{itemize}
        \item Provide guide vane reference $g^*$ to the governor.
        \item Minimize the operation of guide vane opening $g$ to reduce wear and tear.
        \item Optimize the control of guide vane opening $g$ to minimize water hammering and mass oscillation.
        \item Keep the surge tank head $h_{st}$ within its limits and close to the stationary value.
        \item Keep the water flow $q$ above its minimum level.
        \item Optimize the rotational speed of the turbine $\omega$.
    \end{itemize}
    \item Turbine speed control:
    \begin{itemize}
        \item Keep the rotational speed of the turbine $\omega$ within its limits and close to its optimal speed.
        \item Make sure that $\omega$ will recover after a disturbance.
    \end{itemize}
\end{itemize}

Voltage control is another possible task for the MPC, however, it has not been implemented in this paper.

The MPC objective function handles conflicts between the control objectives. For instance, in cases where the output power of the VSHP changes rapidly, fast control of the guide vane opening $g$ is needed to reduce the deviation in turbine rotational speed $\omega$. This will increase the deviation in surge tank head $h_{st}$, increasing mass oscillation and water hammering and thereby increase the cost of the objective function. The MPC compares these costs with the costs of deviation in turbine rotational speed $\omega$ to find the optimal solution. 


\subsection{Model Predictive Controller Dynamic Model}\label{sec:mpcdyn}

The nonlinear MPC dynamic model with its costs and limitations is presented in this section. Except for the modelling of the pressure waves in the penstock $h_p$, it is identical to the model presented in \cite{reigstad2020optimized}. All model parameters are presented in \cite{reigstad2019modelling} and \cite{reigstad2020variable}. The differential equations for the waterway are thereby given as:

\begin{equation}\label{eqww}
    \begin{split}
        \dot{h}_{st} &= \frac{1}{C_s} \left( q_{\mathrm{hr}} -q\right)\\
        \dot{q}_{hr} &= \frac{1}{T_{\mathrm{w2}}} \left( 1 - h_{\mathrm{st}}+f_{0}{\left(q_{\mathrm{hr}}-q \right)}^2-f_{\mathrm{p2}}{q_{\mathrm{hr}}}^2 \right)\\ 
        h &= h_{st} - f_0 \left( q_{\mathrm{hr}} - q  \right)^2  - f_{\mathrm{p1}} q  ^2 + h_{p}\\
        h_{p} &=  - Z_0 \tanh{\left(sT_e\right)} q  = - Z_0 \left( \frac{1-e^{-2T_es}}{1+e^{-2T_es}} \right) q \\
    \end{split}
\end{equation}

where $q_{hr}$ is the head race tunnel flow, $h_{st}$ is the surge tank head, $h$ is the turbine head (pressure difference over the turbine) and $e^{-2T_es}$ is a time delay of $2T_e$.

The turbine model is based on the Euler turbine equation, as presented in \cite{nielsen2015simulation,reigstad2019modelling}.

\begin{equation}\label{eqturb}
    \begin{split}
       P_m &=   \frac{H_{\mathrm{Rt}}}{H_{\mathrm{R}}} \frac{Q_{R}}{Q_{Rt}} \\
        & \quad \left( \left( \frac{\xi q}{g}  \left( \tan{\alpha_{1R}} \sin{ \alpha_1}
         + \cos{\alpha_1} \right) \right) - \psi \omega \right) \frac{q \omega}{h}\\
        \alpha_1 &= \sin^{-1}{ \left(  \frac{Q_{\mathrm{R}}}{Q_{\mathrm{Rt}}} g \sin{\alpha_{1R} } \right)} \\
        \dot{q} &= \frac{1}{T_{\mathrm{w1}}} \left( h \frac{H_{\mathrm{R}}}{H_{\mathrm{Rt}}} -\sigma \left( \omega^2 -1  \right) -  \left( \frac{q }{g} \right)^2 \right) \frac{Q_{Rt}}{Q_{R}}\\
    \end{split}
\end{equation}

The guide vane opening is found from:

\begin{equation}\label{eqgov}
    \begin{split}
        \dot{g} &= \frac{1}{T_{G}} \left( g^* - g \right)
    \end{split}
\end{equation}

while the synchronous generator is modelled by a second-order model:

\begin{equation}\label{eqsg}
    \begin{split}
        \dot{\omega}&= \frac{1}{2H} \left( T_m - P_g/\omega - D \left( \omega^*-\omega \right) \right)\\
    \end{split}
\end{equation}

The virtual synchronous generator (VSG) of the VSHP grid-connected converter controls the output power $P_g$:

\begin{equation}\label{eqvsg}
    \begin{split}
        P_{g} &= i_{g,d} = k_{vsg,p} \Delta f +  k_{vsg,d} \Delta \dot{f}   + P_{g}^{*}\\
        \Delta f &= f-f^*\\
    \end{split}
\end{equation}

The grid frequency is modelled by the swing equation.

\begin{equation}
\begin{split}\label{eqgrid}
    \Delta \dot{f} &= \frac{\omega_s}{2 H_g S_n} \left( P_g + P_{pb} - D_m \Delta  f \right)\\
\end{split}
\end{equation}

where $P_{pb}$ is the power balance of the grid without the VSHP; $P_{pb} = P_{generation} - P_{loads} - P_{losses}$. This variable is estimated from the measured grid frequency $f$ and ROCOF $\dot{f}$ by the PLL.

\begin{equation}\label{eqPe}
\begin{split}
     P_{pb} &= -P_g + \frac{2H_g S_n}{\omega_s} \frac{\omega_{\dot{f}}}{s+\omega_{\dot{f}}} \Delta \dot{f} + D_m \frac{\omega_{f}}{s+\omega_{f}} \Delta f
\end{split}
\end{equation}

The resulting MPC dynamic model DAEs are given in \eqref{eqww}-\eqref{eqgrid} where the states $x$ and inputs $u$ are

\begin{equation}\label{eq:states}
\begin{split}
     x &= [\Delta f  \quad g  \quad q \quad q_{hr} \quad h_{st} \quad \omega]^T\\
    \dot{x} &= [\Delta \dot{ f}  \quad \dot{g}  \quad \dot{q} \quad \dot{q}_{hr} \quad \dot{h}_{st} \quad \dot{\omega}]^T \\
     u &= [P_g^* \quad  g^*]^T
\end{split}
\end{equation}


To optimize the control of guide vane opening $g$ to minimize water hammering, the elastic penstock water column has to be included in the model. One solution is to approximate the classical wave solution by a lumped-parameter equivalent for $\tanh{\left(sT_e\right)}$ as given in \cite{reigstad2019modelling}. However, since the elastic water time constant is low (126 ms), this would require a very short time step of the MPC controller to capture the dynamics of the water hammering. 

Another solution is to utilize that $e^{2T_e s}$ is a time delay of $2T_e$; the time for the pressure wave to move up and down again in the penstock. By setting the time step $\Delta t = 2T_e$, the future pressure waves $h_p$ can be estimated based on the previous pressure waves and flow $q$. From \eqref{eqww}, we get:

\begin{equation}
    \begin{split}
        h_{p,n+1} = - Z_0 \left( q_{n+1} - q_n \right) - h_{p,n}
    \end{split}
\end{equation}

The pressure waves in the penstock $h_p$ affect the turbine head $h$ \eqref{eqww} and thereby the flow $q$. The term $h_p$ is added to the next value of $q$ during the Runge-Kutta method \eqref{rk4a}-\eqref{rk4b}, such that

\begin{equation}
    \begin{split}
        q_{n+1} = q_n &+ \frac{1}{6} \left( k_{1,q} + 2k_{2,q} + 2k_{3,q} + k_{4,q} \right) \\
        &+ \frac{H_R \Delta t}{H_{Rt} T_{w1}} h_{p,n+1} \\
    \end{split}
\end{equation}

The pressure waves also affect the turbine mechanical power $P_m$ \eqref{eqturb} and thereby the turbine rotational speed $\omega$. However, the large inertia $H$ of the turbine and generator will filter the oscillations, and the effect of $h_p$ on $P_m$ is therefore neglected. In addition, the cost for changes in turbine pressure $h$ must be included.


The classical Runge-Kutta method (RK4) is used for numerically integrating the ordinary differential equations. The next value $y_{n+1}$ is found by adding the previous value $y_n$ by a weighted average of four increments ($k_1$-$k_4$) based on the slopes at the beginning, the midpoint and the end of the interval \cite{kincaid2009numerical}.

\begin{equation}\label{rk4a}
    \begin{split}
        y_{n+1} &= y_n + \frac{1}{6} \left( k_1 + 2k_2 + 2k_3 + k_4 \right) \\
        t_{n+1} &= t_n + \Delta t \\
    \end{split}
\end{equation}

for $n = 0, \dots, N-1$, where
 
\begin{equation}\label{rk4b}
    \begin{split}
        k_1 &= \Delta t f \left(t_n, y_n \right) \\
        k_2 &= \Delta t f \left(t_n + \frac{\Delta t}{2}, y_n + \frac{k_1}{2} \right) \\
        k_3 &= \Delta t f \left(t_n + \frac{\Delta t}{2}, y_n + \frac{k_2}{2} \right) \\
        k_4 &= \Delta t f \left(t_n + \Delta t, y_n + k_3 \right) \\
    \end{split}
\end{equation}

\subsection{Costs in MPC Cost Function}\label{ch:costs}

The relative values of the costs determine how the MPC prioritizes between the objectives given in Section \ref{ch:objectives}. A high cost related to an objective causes the MPC controller to prioritize this objective to reduce the cost function. The objectives are prioritized as follows:

\begin{enumerate}
    \item Keep the surge tank head $h_{st}$ within its constraints to avoid damage to the hydraulic system.
    \item Keep the turbine rotational speed $\omega$ within its constraints to avoid undesirable operation conditions of the hydraulic system and damage of the generator.
    \item Minimize water hammering and mass oscillations.
    \item Minimize power oscillations.
    \item Minimize the deviation in the VSHP power reference $P_g^*$ to ensure that the VSHP is contributing to the frequency regulation as intended by the VSG.
    \item Minimize the deviation of the turbine rotational speed $\omega$ from the best efficiency operating point and maximize the turbine efficiency to increase the efficiency of the system.
    \item Keep the water flow $q$ within its constraints to avoid undesired operation conditions of the hydraulic system.
    \item Minimize the deviation in grid frequency $\Delta f$.
\end{enumerate}

The costs are divided into three categories:

\subsubsection{Cost on Deviations in States and Inputs}\label{cof}

The MPC tries to keep the turbine rotational speed $\omega$, the VSHP power reference $P_g^*$  and the VSHP frequency $f$ close to its reference value by considering the cost for deviations in these variables, as given in Table \ref{table_costs}. The variables with the highest corresponding costs will be prioritized. For instance, the cost of deviations in $P_g^*$ is higher than for deviations in turbine rotational speed $\omega$ since $P_g^*$ is not supposed to compensate for deviations in $\omega$ unless $\omega$ is out outside its limits.

\begin{table}[!t]
\renewcommand{\arraystretch}{1.3}
\caption{Cost on deviations in states and inputs}
\label{table_costs}
\centering
\begin{tabularx}{1\linewidth}{|X|cc|}
    \hline
    State & Reference value & Cost factor \\
     &  &  $Q(i,i)$/$R(i,i)$\\
    \hline
    Turbine rotational speed $\omega$& $f(P_{g})$ & 1000\\
    Turbine rotational speed $\omega(N)$& $f(P_{g})$ & 10000\\
    VSHP power reference $P_g^*$ & 0.8 & 1000\\
    VSHP frequency $f$ & $\overline{f}$ & 1e7 \\
    \hline
\end{tabularx}
\end{table}

The VSHP MPC control may contribute to the damping of power oscillations by minimizing the VSHP frequency deviation from the average system frequency, as suggested for HVDC converters in \cite{fuchs2014stabilization}. Since the MPC controls the VSHP power reference $P_g^*$, to achieve this, the cost of deviations in VSHP frequency $f$ is higher than the cost of deviations in VSHP power reference $P_g^*$. The average system frequency $\overline{f}$ found as

\begin{equation}\label{eqOmegaAv}
\begin{split}
     \overline{f} \left( t \right) = \frac{\sum_{i=1}^{n_{gen}} H_i \omega_i \left( t \right) }{\sum_{i} H_i}
\end{split}
\end{equation}

where $H_i$ and $\omega_i$ are, respectively, the generator inertial constant and frequency of the $i^{th}$ generator. A cost for the deviation between the VSHP frequency $f$ and the average system frequency $\overline{f}$ is included in the MPC cost function, as stated in Table \ref{table_costs}. Frequency measurements $\omega_i$ from PMUs for all large generators are needed to calculate the average system frequency $\overline{f}$.

The time constant of the turbine speed dynamics is larger than the time horizon of the MPC. This means that the MPC's last estimated value of the turbine speed does not reach the reference value after a large disturbance. The cost of not reaching the turbine reference speed at the end of the time horizon of the MPC is included by an extra cost for deviation in $\omega (N)$, which is larger than the cost of deviation in turbine speed $\omega$ for each time step.

\subsubsection{Cost on Changes in States and Inputs}

Costs for changes in the pressure waves in the penstock $h_p$ and in guide vane opening reference $g^*$ and $g^*_{-5} $, as given in Table \ref{table_costs_delta}, are introduced to reduce water hammering and mass oscillations. The latter cost also reduces guide vane wear and tear. Since the deviation in $h_p$ is small, the cost factor must be very high to have an effect.
A low value of the cost of change of the guide vane opening reference $g^*$ causes rapid changes in the water flow and mass oscillations. This can be solved by increasing this value; however, the result will be reduced speed of changes of the guide vane opening reference $g^*$ and, thereby, increased deviation in turbine rotational speed $\omega$. The solution is to damp the mass oscillations by adding cost for changes in the guide vane opening reference over a period corresponding to half of the mass oscillation period, in this case, five time steps, such that $g^*_{-5} = g^*_{t} -g^*_{t-5}$.

\begin{table}[!t]
\renewcommand{\arraystretch}{1.3}
\caption{Cost on changes in states and inputs}
\label{table_costs_delta}
\centering
\begin{tabularx}{1\linewidth}{|X|c|}
    \hline
    State & Cost factor $Q_{\Delta t}(i,i)$/$R_{\Delta t}(i,i)$\\
    \hline
    $\Delta g^*$ & 1000 \\
    $\Delta g^*_{-5}$ & 1000 \\
    $\Delta h_p$ & 1e10\\ 
    \hline
\end{tabularx}
\end{table}

\subsubsection{Cost on reduced efficiency in turbine}

The increased cost for allowing the turbine speed to deviate from its optimal value can be added by maximizing the efficiency $\eta$ of the turbine, as given in \eqref{eqeff}. This term in the cost function can be used instead of - or in addition to - the cost of deviation from the optimal turbine speed.

\begin{multline}\label{eqeff}
    \eta = \omega \xi \sqrt{\sigma \left( \omega^2-1 \right)} \\
    \left(  \cos{ \left (\sin^{-1}{\kappa\sin{\alpha_1}}\right)} + \kappa \tan{\alpha_1}  \sin{\alpha_1} \right) -\psi \omega 
\end{multline}

where 
\begin{equation}
\begin{split}
     \kappa &= \frac{Q_R}{Q_{Rt}} g
\end{split}
\end{equation}

The cost-efficiency factor is set to 10 in this case.

\subsection{Constraints and Slack Variables}\label{ch:limits}

The MPC utilizes slack variables to allow the variables to exceed the constraints with an additional cost. This is necessary for obtaining convergence of the optimization problem in some cases. Absolute constraints are given in Table \ref{table_limits}, while the slack variables are presented in Table \ref{table_slack}. 

\begin{table}[!t]
\renewcommand{\arraystretch}{1.3}
\caption{Limits on inputs and variables}
\label{table_limits}
\centering
\begin{tabularx}{1\linewidth}{|X|cc|}
    \hline
    Input & Min. value & Max. value \\
    \hline
    Guide vane opening reference $g^*$ & 0.1 & 1.2 \\
    Converter power $P_g$ & 0 & 1\\
    \hline
\end{tabularx}
\end{table}

\begin{table}[!t]
\renewcommand{\arraystretch}{1.3}
\caption{Slack variables}
\label{table_slack}
\centering
\begin{tabularx}{1\linewidth}{|X|ccc|}
    \hline
    Slack variable & Min. limit & Max. limit & Cost factor $S(i,i)$\\
    \hline
    Water flow $q$ & 0.3 & 1.3 & 1\\
    Surge tank head  $h_{st}$ & 0.5 & - & 1e5\\
    Turbine head $h$ & - & 1.1 & 1e5\\
    Turbine rot. speed $\omega$ & 0.7 & 2 & 1e4\\
    \hline
\end{tabularx}
\end{table}

The guide vane opening reference $g^*$ is limited by the minimum and maximum values of the guide vane opening $g$. The limitation of VSHP output power $P_g$ is set between $0-1 p.u.$ such that the power is delivered to the grid. If the reactive power and the grid voltage are known at the point of common coupling (PCC), the constraints of $P_g$ may be a function of these values to consider the current limit of the converter.
 
The turbine flow $q$ slack variable has a low cost factor since the consequences of exceeding the constraints are low. In contrast, the cost factor of the surge tank head $h_{st}$ slack variable is higher since the consequences of exceeding the constraints are large. Too low surge tank head will cause sand to raise from the sand trap near the surge tank and send it through the turbine, causing increased wear and tear and reduced lifetime of the turbine. The maximum surge tank head is limited by the maximum level of the surge tank to avoid blowout of the surge shaft. However, the maximum constraint on the turbine head $h$ is usually lower, normally 1.1-1.15 p.u., to avoid damage on the turbine blades. In this case, the maximum pressure is limited by the turbine and the maximum constraint of $h$ is set to 1.1 p.u. 

The cost factor related to the turbine rotational speed $\omega$ slack variable is very high since the consequence of exceeding the maximal value is high; generator poles may physically fall off and destroy the generator.
The lower limit of  $\omega$ prevents cases where the turbine produces too low power because of low turbine speed. At low rotational speed and high VSHP output power $P_g$, the electrical torque will be very high. To produce enough mechanical torque to increase $\omega$, the guide vane opening $g$ and turbine flow $q$ must be increased. If $g$, $q$ or the surge tank head $h_{st}$ reach their limits, the MPC controller might not be able to regain the reference turbine speed without reducing the converter output power $P_g$. If both the surge tank head $h_{st}$ and the turbine rotational speed $\omega$ slack variables are activated, the MPC will change the VSHP power reference $P_g^*$ since the cost factor for the VSHP power reference $P_g^*$ deviations is lower than the cost of the slack variables.

\subsection{Reference Turbine Rotational Speed}

For a given turbine flow $q$, and thereby a corresponding stationary VSHP output power $P_{g}$, there exists an optimal turbine rotational speed. Therefore, the turbine rotational speed reference $\omega^*$ is given as a function of the VSHP output power $P_{g}$ \eqref{eqturbopt} to maximize the power production. The function is derived from the hill chart of a reversible pump-turbine presented in \cite{iliev2019variable}. It is provided as a reference value for turbine rotational speed $\omega$ in the MPC cost function, as stated in Table \ref{table_costs}.

\begin{equation}\label{eqturbopt}
    \begin{split}
        0.85 < P_{g} \quad \quad \quad \quad \omega^* &= 1 +0.6 (P_{g}-0.85) \\
        0.73 < P_{g} < 0.85 \quad \omega^* &= 1 + 0.3(P_{g}-0.85) \\
          P_{g} < 0.73 \quad \omega^* &= 0.964 + 0.15(P_{g}-0.73) \\
    \end{split}
\end{equation}

\section{Moving Horizon Estimation}\label{sec:mhe}

MHE is a multivariable estimation algorithm that utilizes a series of measurements and an internal dynamic model of the process to estimate the current states. A major benefit compared to the Kalman filter from \cite{reigstad2020optimized} is the possibility of representation of water hammering in the penstock. This is included in the internal dynamic model presented in Section \ref{sec:mpcdyn}, which is the basis for the MHE. The limits and slack variables from Section \ref{ch:limits} are not included and the cost function is given as: 

\begin{equation}\label{eqcostmhe}
    \begin{split}
        J_{N_{mhe}} \left( x,u \right) &= \sum_{i=k-N_{mhe}}^{k} \left\| \tilde{y}(i) - y(i) \right\|_V^2 \\
        & \quad + \sum_{i=k-N_{mhe}}^{k-1} \left\| \tilde{u}(i) - u(i) \right\|_W^2
    \end{split}
\end{equation}

subjected to

\begin{equation}
    \begin{split}
        x_{t+1} &= g\left( x_t, u_t \right) \\
        x_0, u_{-1} &= \text{given} \\
    \end{split}
\end{equation}

where 

\begin{equation}\label{eqcostmhe3}
    \begin{split}
        y &= [\Delta f  \quad g   \quad h_{st} \quad \omega \quad h \quad P_m \quad P_g]^T\\
        u &= [P_g^* \quad  P_{pb} \quad  g^*]^T
    \end{split}
\end{equation}

The variables $y$ and $u$ are, respectively, the system outputs and inputs while $\tilde{y}$ and $\tilde{u}$ are the system output and inputs of the estimated model. The relative cost of deviation between the estimated and measured system outputs and system inputs are found from the standard deviation in Gaussian noise such that:

\begin{equation}
    \begin{split}
       V &=  \begin{bmatrix}
            \sigma_{y(1)} &  0 &  0 \\
            0 & \ddots & 0 \\
            0 &  0 & \sigma_{y(n_y)} 
        \end{bmatrix}^{-1}\\
        &= \diag{50,100,100,1000,100,1,1}\\
    \end{split}
\end{equation}

\begin{equation}
    \begin{split}
       W &= \begin{bmatrix}
            \sigma_{u(1)} &      0 &  0 \\
            0 & \ddots & 0 \\
            0 &  0 & \sigma_{u(m)} \\
        \end{bmatrix}^{-1}\\
        &= \diag{100,1000,10}\\
    \end{split}
\end{equation}

The initial and previous values for the states and for the pressure waves $h_p$ are found by the MHE. The initial value of $h_p$ is therefore based on the previous measurements, primarily of $h$ and $h_{st}$.

\section{Results and Discussion}\label{Results}

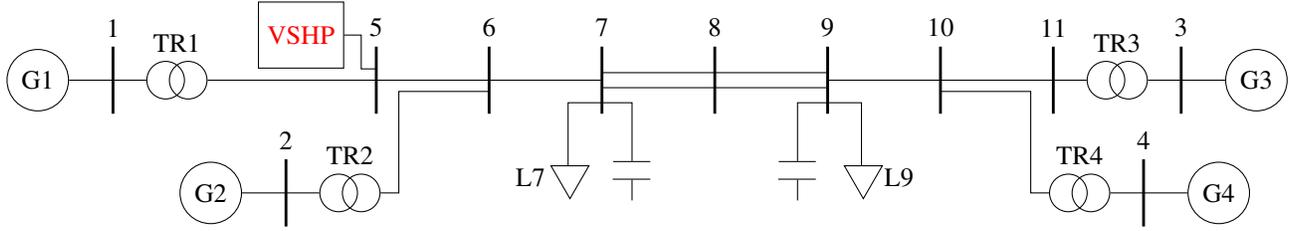
\begin{figure*}[!t]
\centering
\begin{tikzpicture}[auto, node distance=1.5cm,>=latex']
    \node [blockbus] (bus1) {};
    \node [generator, left of = bus1, node distance =1.0cm] (gen1) {G1};
    \draw [-] (gen1) -- (bus1);
    \node [tr, right of = bus1, node distance =0.7cm] (tr1a) {};
    \node [tr, right of = tr1a, node distance =0.3cm] (tr1b) {};
    \draw [-] (bus1) -- (tr1a);
    \node [blockbus, right of = tr1b, node distance = 25cm] (bus5) {};
    \draw [-] (tr1b) -- (bus5);
    \node [blockbus, right of = bus5, node distance = 15cm] (bus6) {};
    \draw [-] (bus5) -- (bus6);
    \node [blockbus, right of = bus6, node distance = 15cm] (bus7) {};
    \draw [-] ([yshift=-0.0cm]bus6.east) -- ([yshift=-0.0cm]bus7.west);
    \node [blockbus, right of = bus7, node distance = 15cm] (bus8) {};
    \draw [-] ([yshift=0.1cm]bus7.east) -- ([yshift=0.1cm]bus8.west);
    \draw [-] ([yshift=-0.1cm]bus7.east) -- ([yshift=-0.1cm]bus8.west);
    \node [blockbus, right of = bus8, node distance = 15cm] (bus9) {};
    \draw [-] ([yshift=0.1cm]bus8.east) -- ([yshift=0.1cm]bus9.west);
    \draw [-] ([yshift=-0.1cm]bus8.east) -- ([yshift=-0.1cm]bus9.west);
    \node [blockbus, right of = bus9, node distance = 15cm] (bus10) {};
    \draw [-] ([yshift=-0.0cm]bus9.east) -- ([yshift=-0.0cm]bus10.west);
    \node [blockbus, right of = bus10, node distance = 15cm] (bus11) {};
    \draw [-] (bus10) -- (bus11);
    \node [tr, right of = bus11, node distance =0.7cm] (tr3a) {};
    \node [tr, right of = tr3a, node distance =0.3cm] (tr3b) {};
    \draw [-] (bus11) -- (tr3a);
    \node [blockbus, right of = tr3b, node distance = 7cm] (bus3) {};
    \draw [-] (tr3b) -- (bus3);
    \node [generator, right of = bus3, node distance =1.0cm, node distance =1.0cm] (gen3) {G3};
    \draw [-] (bus3) -- (gen3);
    
    \node [point, right of = tr1b, node distance =2.3cm] (tr1br) {};
    \node [tr, below of = tr1br, node distance =1.5cm] (tr2a) {};
    \node [tr, left of = tr2a, node distance =0.3cm] (tr2b) {};
    \node [point, right of = tr2a, node distance =0.5cm] (tr2ar) {};
    \draw [-] ([yshift=-0.15cm]bus6.west) -| (tr2ar) -- (tr2a);
    
    \node [blockbus, left of = tr2b, node distance = 7cm] (bus2) {};
    \draw [-] (tr2b) -- (bus2);
    \node [generator, left of = bus2, node distance =1.0cm] (gen2) {G2};
    \draw [-] (bus2) -- (gen2);
    
    \node [point, left of = tr3a, node distance =0.5cm] (tr3al) {};
    \node [tr, below of = tr3al, node distance =1.5cm] (tr4a) {};
    \node [tr, right of = tr4a, node distance =0.3cm] (tr4b) {};
    \node [point, left of = tr4a, node distance =0.5cm] (tr4al) {};
    \draw [-] ([yshift=-0.15cm]bus10.east) -| (tr4al) -- (tr4a);
    
    \node [blockbus, right of = tr4b, node distance = 7cm] (bus4) {};
    \draw [-] (tr4b) -- (bus4);
    \node [generator, right of = bus4, node distance =1.0cm] (gen4) {G4};
    \draw [-] (bus4) -- (gen4);
    
    \node [point, left of = bus7, node distance =1.2cm] (bus7l) {};
    \node [load, below of = bus7l] (load7) {};
    \draw [-] ([yshift=-0.3cm]bus7.east) -| (load7);
    
    \node [point, left of = bus9, node distance =0.3cm] (bus9r) {};
    \node [load, below of = bus9r] (load9) {};
    \draw [-] ([yshift=-0.3cm]bus9.west) -| (load9);
    
    \node [point, right of = bus7, node distance =0.4cm] (bus7r) {};
    \node [capacitor, below of = bus7r,  node distance =1.2cm] (cap7) {};
    \draw [-] ([yshift=-0.3cm]bus7.west) -| (cap7);
    \node [point, below of = cap7, node distance =0.4cm] (cap7b) {};
    \draw [-] (cap7) -- (cap7b);
    
    \node [point, left of = bus9, node distance =0.4cm] (bus9l) {};
    \node [capacitor, below of = bus9l,  node distance =1.2cm] (cap9) {};
    \draw [-] ([yshift=-0.3cm]bus9.east) -| (cap9);
    \node [point, below of = cap9, node distance =0.4cm] (cap9b) {};
    \draw [-] (cap9) -- (cap9b);

    \node [align=center] at ($(bus1)+(0,0.7)$) {1};
    \node [align=center] at ($(bus2)+(0,0.7)$) {2};
    \node [align=center] at ($(bus3)+(0,0.7)$) {3};
    \node [align=center] at ($(bus4)+(0,0.7)$) {4};
    \node [align=center] at ($(bus5)+(0,0.7)$) {5};
    \node [align=center] at ($(bus6)+(0,0.7)$) {6};
    \node [align=center] at ($(bus7)+(0,0.7)$) {7};
    \node [align=center] at ($(bus8)+(0,0.7)$) {8};
    \node [align=center] at ($(bus9)+(0,0.7)$) {9};
    \node [align=center] at ($(bus10)+(0,0.7)$) {10};
    \node [align=center] at ($(bus11)+(0,0.7)$) {11};
    \node [align=center] at ($(load7)+(-0.5,0)$) {L7};
    \node [align=center] at ($(load9)+(0.5,0)$) {L9};
    \node [align=center] at ($(tr1a)+(0.15,0.5)$) {TR1};
    \node [align=center] at ($(tr2a)+(-0.15,0.5)$) {TR2};
    \node [align=center] at ($(tr3a)+(0.15,0.5)$) {TR3};
    \node [align=center] at ($(tr4a)+(0.15,0.5)$) {TR4};
    
    \node [point, left of = bus5, node distance =1.0cm] (bus5l) {};
    \node [block, above of = bus5l, node distance = 0.6cm, text=red,] (vshp) {VSHP};
    \node [point, right of = vshp, node distance =0.8cm] (vshpr) {};
    \draw [-] (vshp) -- (vshpr) |- ([yshift=0.15cm]bus5.west);
    
\end{tikzpicture}
\caption{Kundur two-area system} \label{figKun}
\end{figure*}

This section presents the results of the dynamic simulation with the non-linear MPC controller and compares it with the linear MPC controller developed in \cite{reigstad2020optimized}. The dynamic performance of the MPC controller is tested on the grid presented in \cite{reigstad2019variable} and Figure \ref{figKun}, based on the Kundur two-area system. The hydraulic system, the synchronous generator and the converters are modelled as presented in \cite{reigstad2019variable} with some modifications: 

\begin{itemize}
    \item The active power control of the grid-connected converter is replaced by a VSG, as presented in \cite{reigstad2020variable}.
    \item The VSG power reference is provided by the MPC.
    \item The governor control is replaced by the guide vane reference from the MPC.
\end{itemize}

An overall scheme of the system is shown in Figure \ref{figvshp}.

Cases with both overproduction and underproduction are investigated by first reducing the load by 160 MVA at Bus 7 at time $t=0s$ and thereby increasing the load back to the initial value at $t=60s$. 

\subsection{Performance of Moving Horizon Estimation}

\begin{figure}[!t]
\centering
    \includegraphics[scale=0.8]{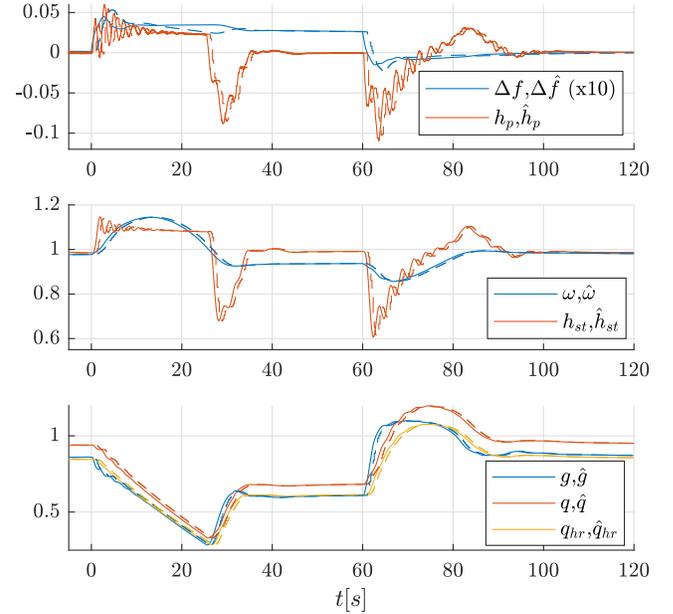}
    \caption{Comparison of real states (solid) and estimated states by the MHE (dashed): Frequency deviation $\Delta f$, penstock pressure waves $h_p$, turbine rotational speed $\omega$, surge tank head $h_{st}$, guide vane opening $g$, penstock flow $q$ and head race tunnel flow $q_{hr}$. } \label{fig606}
\end{figure}

The MPC controller needs reliable estimations of the states to find an optimal solution to the control problem. The MHE utilizes the same dynamic model as the MPC and the 10 previous measurements of the model inputs and outputs to estimate the current state. The deviation between the real values of the states (solid) and the estimated states (dashed) are shown in Figure \ref{fig606}.

For most of the states, the estimated values are following the real values with a time delay of approximately 1 second. This time delay causes the estimations of the surge tank head $h_{st}$ and the penstock pressure waves $h_p$ to be in anti-phase with the real values,

\subsection{Non-linear MPC compared to linear MPC}

\begin{figure}[!t]
\centering
    \includegraphics[scale=0.8]{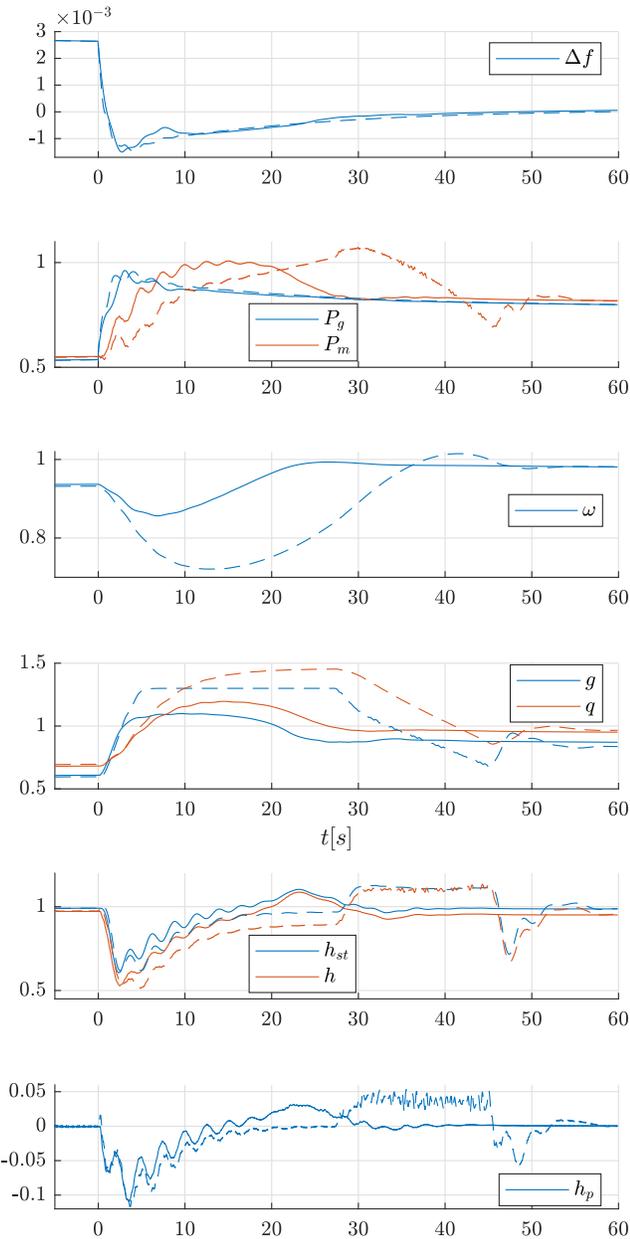}
    \caption{Comparison of non-linear MPC (solid) as presented in this paper and linear MPC as presented in \cite{reigstad2020optimized} (dashed): Frequency deviation $\Delta f$, grid converter power $P_g$, turbine mechanical power $P_m$, turbine rotational speed $\omega$, guide vane opening $g$ and penstock flow $q$, surge tank head $h_{st}$, turbine head $h$, penstock pressure waves $h_p$.} \label{fig608}
\end{figure}

Figure \ref{fig608} compares the dynamic results from the non-linear MPC controller presented in this paper (solid) with the linear MPC controller presented in \cite{reigstad2020optimized} (dashed). The non-linear MPC outperforms the linear MPC controller in most regards. The most recognizable and important improvement is the reduction of the deviation in turbine rotational speed $\omega$. The non-linear MPC responds slightly faster by increasing the guide vane opening $g$ after the disturbance. The linear MPC is slower, causing higher deviations in turbine rotational speed $\omega$ and thereby a self-energizing effect due to reduced mechanical power $P_m$. Besides, the guide vane opening is increased too much because the linearization of the turbine model causes an inaccurate prediction of the mechanical power of the turbine. The prediction of the mechanical power $P_m$ for the non-linear MPC is better, reducing the overshoots in turbine rotational speed $\omega$, gate opening $g$ and penstock flow $q$ after a disturbance.


\subsection{Effect of Damping of Power Oscillation Damping}

\begin{figure}[!t]
\centering
    \includegraphics[scale=0.8]{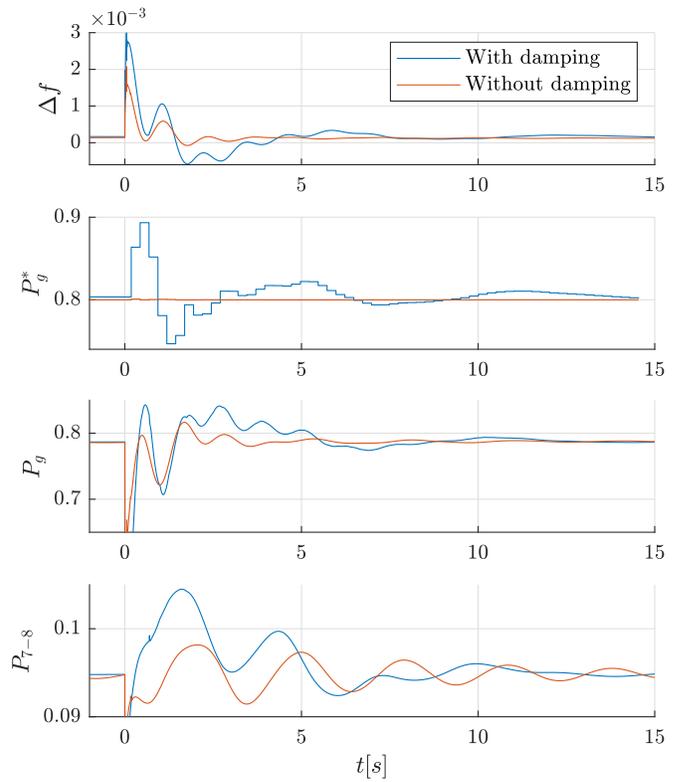}
    \caption{Effect of power oscillation damping by the MPC: Frequency deviation $\Delta f$, grid converter power reference $P_g^*$, grid converter power $P_g^*$ and active power between the two areas of the grid $P_{7-8}$.} \label{fig607}
\end{figure}

The non-linear MPC includes the damping of power oscillations, as explained in Section \ref{cof}. Figure \ref{fig607} shows how this function affects the frequency deviation $\Delta f$, the grid converter power reference $P_g^*$, the grid converter power $P_g$ and the active power between the two areas of the grid $P_{7-8}$ after a $50ms$ three-phase short-circuit at Bus 8 at $t=0$. When the damping function is activated, the MPC adjusts the VSHP output power reference $P_g^*$ to minimize the deviation between the local frequency at the VSHP and the average system frequency $\overline{f}$ and thereby damp the power oscillations. The VSHP output power reference $P_g^*$ is first increased to increase the local frequency before it is reduced. The effect is seen in power between the two areas $P_{7-8}$. The magnitudes of the oscillations are similar with and without the damping function for the first 5 seconds. Subsequently, the damping of the dominated mode is improved significantly by the damping function, as seen in the figure. 

\subsection{Effect of Modelling Water Hammering in the Penstock}

\begin{figure}[!t]
\centering
    \includegraphics[scale=0.8]{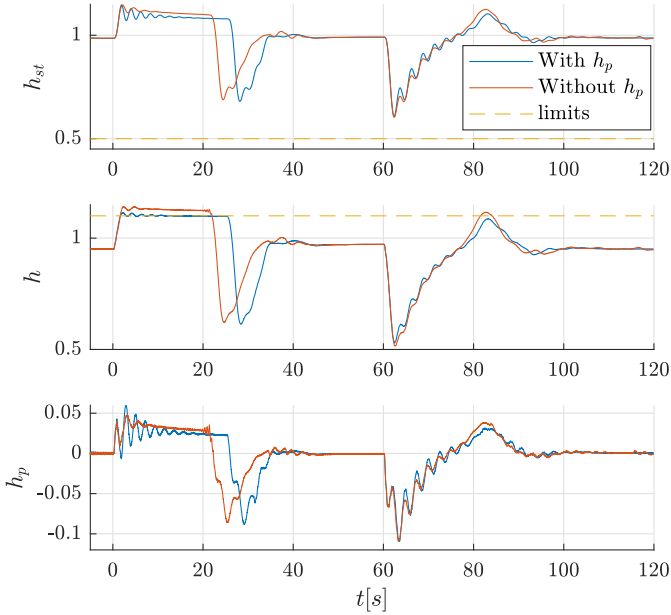}
    \caption{Effect of modelling water hammering in the penstock: Surge tank head $h_{st}$, turbine head $h$ and penstock pressure waves $h_p$.} \label{fig609}
\end{figure}

The non-linear MPC also includes modelling of the water hammering in the penstock. The effect of this is shown in Figure \ref{fig609} where the presented non-linear MPC is compared with a version of the MPC not including the modelling of the water hammering. If the water hammering is not modelled, the turbine head $h$ will exceed its limits when the gate is closing at maximal speed. The gate closing speed will be faster, resulting in less deviation in turbine rotational speed. The difference between the two cases will be less when the gate is opening since the turbine head $h$ or surge tank head $h_{st}$ is not a constraint in this case. 

The oscillations in the penstock pressure waves are larger from 0-20 seconds if they are included in the MPC model than if they are not. The reason for this is that the MPC tries to keep the turbine head $h$ at its maximum value by counteracting the oscillations in the penstock caused by mass oscillations. The guide vane opening is slightly adjusted to obtain this, causing oscillations in the penstock with the same frequency as the mass oscillation.

\section{Conclusion}\label{Conclusion}

With the increased share of variable energy production, such as wind and solar, the demand for flexible generation and consumption is increasing. Most producers have limited energy storage and are not able to increase production quickly. Besides, a fast reduction in power will normally cause increased energy losses. The advantage of a variable speed hydropower (VSHP) plant is the possibility to utilize the energy storage in the rotation masses, making it able to both increase and decrease its output power almost instantaneously. It is, therefore, suitable for delivering both virtual inertia and fast frequency regulation. This paper has described the development of a model predictive controller (MPC) to coordinate the control of the hydro turbine and the VSHP frequency converter, and at the same time considering the constraints in the electric and the hydraulic systems. A moving horizon estimator (MHE) is applied for the estimation of the state variables.

The non-linear MPC presented in this paper shows improved performance compared to an earlier investigated linear MPC. The improvement is primarily due to a more accurate calculation of turbine power, causing less overshoot in the turbine speed after a disturbance. Furthermore, the proposed modelling of water hammering in the penstock improves the calculation of turbine pressure and thereby ensures that maximum pressure is not exceeded. The results do also show that the MPC can contribute to damping power oscillations by adding a cost for the deviation between the local and average frequency to the optimization function. 

With the promising simulation results, the controllers need to be implemented and tested in a laboratory for further verification of performance. Future work will include real-time hardware-in-the-loop, the time delay of the controller and signal processing and a more realistic model of the hydraulic system. 


%





\ifCLASSOPTIONcaptionsoff
  \newpage
\fi



\bibliographystyle{IEEEtran}
\bibliography{Paper6}
\end{document}